\documentclass[manuscript]{aastex}

\usepackage{natbib}
\usepackage{graphicx}
\usepackage{amsmath}
\usepackage{epsfig}

\shorttitle{Water and ammonia in L1157-B1} \shortauthors{Viti et al.}

\begin{document}
   \title{L1157-B1: Water and ammonia as diagnostics of shock temperature}

   \author{
Viti S. \altaffilmark{1},
Jim\`enez-Serra, I \altaffilmark{2},
Yates J. A. \altaffilmark{1},
Codella C. \altaffilmark{3},
Vasta M. \altaffilmark{3},
Caselli P. \altaffilmark{4,3}
Lefloch B. \altaffilmark{5,6}
Ceccarelli C. \altaffilmark{5}}
\email{sv@star.ucl.ac.uk}

   \altaffiltext{1}
{Department of Physics and Astronomy, University College London, Gower Street, London, WC1E 6BT, UK}
\altaffiltext{2}{Harvard-Smithsonian Center for Astrophysics, 60 Garden Street, Cambridge, MA 02138, USA }
\altaffiltext{3}{INAF, Osservatorio Astrofisico di Arcetri, Largo E. Fermi 5, 50125, Firenze, Italy}
\altaffiltext{4}{School of Physics and Astronomy, University of Leeds, Leeds, LS2 9JT, UK}
\altaffiltext{5}{UJF-Grenoble 1/CNRS-INSU, Institut de Planétologie et d?Astrophysique de Grenoble (IPAG) UMR 5274, Grenoble, France}
\altaffiltext{6}{Centro de Astrobiologia, CSIC-INTA, Carretera de Ajalvir, Km 4, Torrejon de Ardoz, 28850, Madrid, Spain}

\begin{abstract}
We investigate the origin and nature of the profiles of water and ammonia observed toward the 
L1157-B1 clump as part of the HIFI CHESS survey (Ceccarelli et al. 2010) using a new code coupling a gas-grain chemical model with a parametric shock model. First results from the unbiased survey (Lefloch et al. 2010, Codella et al. 2010) reveal different molecular components at different excitation conditions coexisting in the B1 bow shock structure, with NH$_3$, H$_2$CO and CH$_3$OH emitting only at relatively low outflow velocities whereas H$_2$O shows bright emission at high velocities. Our model suggests that these differences are purely chemical and can be explained 
by the presence of a C-type shock whose maximum temperature must be close to 4000 K along the B1 clump. 
\end{abstract}
 


\section{Introduction}
During the earliest stage of star formation, protostars
generate fast jets.
The supersonic
impact between the jet and the parent cloud generates shock fronts
propagating through the high-density gas surrounding the protostar.
The shocks heat and compress the ambient gas, and
complex chemistry driven by endothermic chemical reactions,
ice grain mantle sublimation and sputtering occur.
Several molecules, such as H$_2$O, NH$_3$, CH$_3$OH, H$_2$CO, undergo abundance 
enhancements by orders of magnitude (e.g. 
Draine et al. 1983;  Kaufman \& Neufeld, 1996),
as observed at mm-wavelengths in a number of outflows (Garay et al. 1998;
Bachiller \& P\'erez Guti\'errez 1997; Jim\'enez-Serra et al. 2005; J\o{}rgensen et al. 2007).

One of the best laboratories is the L1157 region located at
a distance between 250 and 440 pc, where a Class 0 protostar (L1157-mm) drives a
chemically rich bipolar outflow (e.g, Gueth et al. 1996, Bachiller et al. 2001).
The outflow is associated with several bow
shocks seen in IR H$_2$ (Nisini et al. 2010), in CO (Gueth
et al. 1996) as well as in other molecules (Bachiller \& P\'erez Guti\'errez 1997)
thought to be released from icy mantles such as
H$_2$CO, CH$_3$OH, and typical tracers of
high-speed shocks such as SiO, sputtered from dust grains (Caselli et al. 1997; Gusdorf et al. 2008; Jim\'enez-Serra et al. 2008).
The brightest bow-shock, L1157-B1, has been extensively
studied from the ground using IRAM-PdB and NRAO-VLA interferometers,
revealing a rich and clumpy structure
(Tafalla \& Bachiller 1995; Benedettini et al. 2007; Codella et
al. 2009).

As part of the Herschel Key Program CHESS
(Chemical Herschel Surveys of Star forming regions, Ceccarelli et al. 2010),
L1157-B1 has been investigated with a spectral survey in the
$\sim$500--2000 GHz interval using the HIFI instrument
to provide high-resolution (1 MHz) spectra.
Preliminary results (Codella et al. 2010; Lefloch et al. 2010) have not only confirmed the chemical
richness of the L1157-mm outflow, but have also revealed the presence of
different molecular components at different excitation conditions
coexisting in the B1 structure.
In addition, the CHESS data have allowed, for the first time,
comparison of the line profiles of tracers of shocked material such as CH$_3$OH, H$_2$CO,
with those due to the fundamental transitions of water and ammonia.
These species show two different kinds of profiles,
with CH$_3$OH, H$_2$CO,
and NH$_3$ emitting only at relatively low outflow velocities ($\sim$15 km s$^{-1}$
with respect to the ambient velocity), whereas
H$_2$O shows bright emission even at the highest velocities (up to
30 km s$^{-1}$), as seen in SiO (e.g. Zhang et al. 1995) and CO (Lefloch et al. 2010).
The intensity of methanol, formaldehyde, and ammonia 
with respect to H$_2$O decreases
towards higher outflow velocities suggesting, in case of optically thin
emission, a similar decrease in the abundance ratios.
Note that while SiO and CO are not products of surface reactions, H$_2$O, NH$_3$, CH$_3$OH and H$_2$CO are all highly enhanced on the dust grains, and therefore the distinct behaviour of water is worth investigating.

High NH$_3$ abundances in the gas phase have
been observed in many warm regions (e.g. hot cores and outflows, Beuther et al. 2005; Sepulveda et al. 2011).
NH$_3$ is believed to be mainly a direct product of grain surface reactions (although at high 
temperatures
ammonia can also form in gas phase, reaching abundances of $\sim$ 10$^{-5}$ - see Section 3)
with nitrogen hydrogenating until saturation as it sticks to the dust
during the cold phase (e.g. Brown et al. 1988; Viti et al. 2004a). 
Water is enhanced by the release of the icy mantles {\it as well as} by
endothermic reactions occurring in the warm ($\geq 220$ K) shocked gas, which convert the bulk of atomic
oxygen into water (e.g. Kaufman \& Neufeld 1996, and references therein).

In this Letter we investigate the differences between the water and ammonia profiles by using a new code
which couples a gas-grain chemical model with a parametric shock model.
 
\section{The model}

The gas-grain chemical code UCL\_CHEM (Viti et al. 2004a) was coupled
with the parametric shock model developed by Jim\'enez-Serra et
al. (2008). UCL\_CHEM is a time dependent gas-grain chemical model in two phases.
During Phase I, gravitational collapse,
gas-phase
chemistry and sticking on to dust particles with subsequent processing
occur. This phase simulates the
formation of high density clumps and starts from a 
diffuse ($\sim$ 100 cm$^{-3}$) medium in atomic form (apart from a
fraction of hydrogen in H$_2$). The initial abundances adopted are
solar but different initial values for the sulfur elemental
abundance were also used. Atoms and molecules are
depleted onto grain surfaces as in Rawlings et al. (1992, Eq. 2) 
and hydrogenate when possible. The
depletion efficiency is determined by the fraction of the gas-phase
material that is frozen onto the grains. This approach allows a
derivation of the ice composition. 
Phase II
computes the time dependent chemical evolution of the gas
and dust once the clump has formed and stellar activity is present (in
the form of a protostar and/or outflows). Full details of the code can
be found in Viti et al. (2004a). For this work the UCL\_CHEM was
coupled with the parametric shock model of Jim\'enez-Serra et
al. (2008) which
calculates the
physical structure of a plane-parallel steady-state C-shock that
propagates with a velocity $v_s$ through an unperturbed medium. This approximation was validated by comparing its
results with those from magnetohydrodynamic (MHD)
C-shock modelling (Flower et al. 2003, Kaufman \& Neufeld 1996). 
Although not strictly accurate,
the general behaviour of
the velocity of the ion and neutral fluids 
(i.e. $v_i$ and $v_n$), of the density
of the neutral particles ($n_n$), and of the temperature of the ions and neutrals
($T_i$ and $T_n$), mimics the MHD physical structure of C-shocks
(Figures 1 and 2 in Jim\'enez-Serra et al. 2008).

In addition to the C-shock physical structure, the parametric model includes the sputtering of dust grains following the formalism used by Caselli et al. (1997), updated with the results from May et al. (2000). In our chemical
model, the icy mantles are sputtered
once the dynamical age
across the C-shock has reached the `saturation time-scales'
(or $t_{sat}$; Jim\'enez-Serra
et al. 2008). These are the time-scales for which
almost all molecular material within the mantles is released into the
gas.
In Jim\'enez-Serra et al. (2008), $t_{sat}$ were
originally derived for SiO (see their Table 5). However, 
we adopt the same $t_{sat}$ values for H$_2$O and
NH$_3$ because the mathematical expressions of the fraction of SiO, H$_2$O
and NH$_3$ sputtered from the grain mantles are the same for all these
molecular species except for their initial abundance contained in the
mantles (see Appendix B in Jim\'enez-Serra et al. 2008). The adopted
$t_{sat}$ for H$_2$O and NH$_3$ were finally scaled for the different
initial H$_2$ densities explored in our study, by
assuming \footnote{The cooling time-scales vary
roughly as $n_i^{-1}$, where $n_i$ is the density of the ion fluid and is
proportional to the H$_2$ density of the gas - see e.g. Chi\`eze et al. 1998; Lesaffre et al. 2004).} that $t_{sat}$$\propto$$n_n^{-1}$.
\par
For this work, Phase I
was used to form the pre-existing (to the shock) clump, B1. We do not
know how dense the pre-existing clump was; however modelling of the clumps
along a similar outflow (CB3; Viti et al. 2004b; Benedettini et
al. 2006) has shown that in order for the chemistry to be as rich as
observed, some level of clumpiness must pre-exist the shock; hence we did
not consider final densities for Phase I lower than 1000 cm$^{-3}$
(see Table 1). 
The shock
model was then included in the Phase II of the chemical model.  

\par A total of 15
models were run (Table 1) where we varied: i)
the final densities of Phase I (and hence the initial density of the
clump before it became shocked); we have restricted these densities to the range between
10$^3$ to 10$^5$ cm$^{-3}$, 
similar to those derived by G\"usdorf et al. (2008) and Nisini et al.
(2010) for the B1 clump. Higher pre-shock densities have not been
considered in this study because the collisional dissociation of H$_2$ would
highly overcome the formation rate of H$_2$, leading to the development of a
J-type discontinuity within the physical structure of the shock for the
shock velocities we are considering (Le Bourlot et al. 2002). However, we
note that, although most of the shocks along the L1157-mm outflow are
non-dissociative (or C-type; see Nisini et al. 2010), we cannot rule out
the possibility that a compact J-type shock component is present in clump
B1, as suggested by the detection of [SiII], [FeII] or [SI] emission
arising from this condensation (Neufeld et al. 2009).
The final density and initial velocity also determine the maximum temperature
attainable by the neutral gas within the shock (Draine et al. 1983 - see Figures 8b and 9b for the assumed values of T$_{n, max}$ shown in Table 1; Jim\'enez-Serra et al. 2008).  
Flower \& Pineau des For\^ets (2003) predict higher temperatures for the same density 
and initial velocities at the beginning of the shock than Jim\'enez-Serra et al. (2008), 
although the values converge in the middle of the dissipation region. 
Since, however, recent MHD calculations from van Loo et al. (2009) 
seem to agree with the lower estimates, we adopt the calculations from Jim\'enez-Serra et al. (2008);
ii) the initial elemental sulfur
abundance: we use 1, 1/10 and 1/100 of the solar value; (iii) the degree of depletion, from 1 to 60\% of gas depleted by the end of Phase I, an arbitrary range;
(iv) the shock velocity from
35 to 60 kms$^{-1}$.
{\small
\begin{table*}
  \caption{Model parameters: Model number, pre-shock density, shock velocity, saturation time, maximum temperature of the neutral gas, degree of depletion 
at the end of Phase I, and initial sulfur abundance (where 1 stands for solar).}
    \begin{tabular}{lllllll}
\hline
Model & n(H$_2$) & Vs & T$_{sat}$ & T$_{n,max}$ & Depl. & S\\
\hline
& cm$^{-3}$ &  km s$^{-1}$ & yrs & K  & &  \\
\hline
1 & 10$^5$ & 40 &4.6 & 4000 & 1\% & 1 \\
2 & 10$^5$ & 40 &4.6 & 4000 & 15\% & 1 \\
3 & 10$^5$ & 40 &4.6 & 4000 & 30\%  & 1\\
4 & 10$^5$ & 35 &4.6 & 3200 & 30\% & 1 \\
5 & 10$^5$ & 35 &4.6&3200 & 1\% & 1 \\
6 & 10$^3$ & 40 & 455 &2200 & 10\% & 1 \\
7 & 10$^4$ & 40 & 45.5 & 2200 & 15\%  & 1\\
8 & 10$^4$ & 60 & 38& 4000 & 30\% & 1 \\
9 & 10$^3$ & 60 & 380 &4000 & 30\% & 1 \\
10 & 10$^3$ & 60 & 380 &4000 & 10\% & 1 \\
11 & 10$^5$ & 40 &4.6 & 4000 & 30\%  & 1/10 \\
12 & 10$^5$ & 40 &4.6 & 4000 & 30\%  & 1/100 \\
13 & 10$^5$ & 40 &4.6 & 4000 & 60\% & 1 \\
14 & 10$^4$ & 40 & 45.5 & 2200 & 60\% & 1 \\
15 & 10$^5$ & 35 &4.6 & 3200 & 60\% & 1 \\
\hline
\end{tabular}
\end{table*}
}
\par
Finally as a first attempt at reproducing the {\it trends} of the molecular line profiles,
we input the fractional abundances, densities and temperatures as a function of distance and velocity 
produced by the chemical and shock model 
to the radiative transfer model SMMOL (Rawlings \& Yates
2001), along with values for the dust temperature and radiation field
intensity. The SMMOL code has accelerated $\lambda$-iteration, a numerical scheme that
solves the exact radiative transfer problem in multilevel non-local
conditions, using an approximate lambda operator to derive `pre-conditioned' statistical-equilibrium equations (e.g. Rybicki and Hummer 1991). 
We ran SMMOL for
o-H$_2$O, and NH$_3$. 
The collisional data for water and ammonia were taken from 
Faure et al. (2007) and Danby al. (1988) respectively. New 
collisional data for ammonia are available (Maret et al. 2009); however as they only cover up to temperatures of 100K and para-H$_2$, we have, for this work, used the older data.
While water had been modelled by
SMMOL before (Lerate et al. 2010) this is the first time that
SMMOL was used to model the line intensities and profiles of ammonia.
This is the first attempt at using a non LTE non LVG radiative transfer code to model the high resolution 
H$_2$O and NH$_3$ lines observed
by HIFI. 

\section{Results} 
The aim of this work is 
to understand qualitatively the origin and history of the molecular gas.
Codella et al. (2010) find that the
NH$_3$/H$_2$O intensity ratio decreases at high
velocities implying a similar decrease in the abundance ratios. This
is the first statement that we will test: 
Figure
1 shows the chemical evolution of selected species as a function of
distance within the shock ($z$ in units of cm) for four of our models (Model 1, 3, 4 and
7), chosen as examples of variations in depletion, pre-shock densities and shock velocities. 
We note that the value for the initial abundance of sulfur does 
not affect the water and ammonia behaviour at all and we won't 
discuss it further. An increase in depletion efficiency has the effect of increasing all the species that are formed or enhanced on the dust surfaces, but does not affect the abundance trends. Varying the pre-shock density and the shock velocity (and hence the neutral gas shock temperature) has a profound effect on the abundances trends.
For Models 1 and 3 water behaves differently from the other three species in
that it always remains abundant after the gas cools down, i.e it
remains abundant at higher velocities within the post-shock gas 
(v$_n$ $>$ 15 km s$^{-1}$). At first glance, this is
certainly consistent with the assumption proposed by Codella et al. (2010) 
that the NH$_3$/H$_2$O decrease in intensity at high velocities corresponds to a decrease in their abundance ratio.
In Models 4 and 7, on the other hand, ammonia remains high in abundance
at high velocities. The key difference among these models is the maximum
temperature reached by the gas due to the passage of the shock, and this, in turn, constrains the minimum pre-shock density for the clump (for models 
with the same shock velocities). Indeed, if v$_s$ = 40 km s$^{-1}$ only models with initial H$_2$ densities greater or equal to 10$^5$ cm$^{-3}$ show temperatures of the neutral fluid of $\sim$ 4000 K in the postshock gas, reproducing the observed differences in the NH$_3$/H$_2$O abundance ratio.
The fact that Model 7, where the 
initial density of the B1 clump is only 10$^4$ cm$^{-3}$, does not fit
the observed behaviour is consistent with previous models
of chemically rich clumps along outflows (Viti et al. 2004b), and supports the idea of an inhomogeneous medium with dense pre-existing clumps along L1157. 
Finally we also note that the observed
decrease in the H$_2$CO/H$_2$O and CH$_3$OH/H$_2$O abundance ratios with increasing
velocity within the shock (Codella et al. 2010), is better reproduced by
Models 1 and 3 than by Models 4 and 7.

\par
For ammonia to be abundant in the gas phase alone, high temperatures are required; 
by running several UCL\_CHEM Phase I test models (with constant densities of 10$^5$ cm$^{-3}$, 
with and without freeze out) we find that high abundances (at least 10$^{-6}$) of pure gas phase ammonia 
are only reached if: (i) the temperature is higher than 800 K (as its main route of formation, H$_2$ + NH$_2$, is endothermic, with a barrier of $\sim$ 1400K); 
(ii) no substantial freeze out occurs; (iii) the 
density of the gas is high (in our case 10$^5$ cm$^{-3}$) for at least 10$^4$ years. These results
are of course consistent with previous findings (see review by Ho \& Townes 1983).
Taking as an example Model 3,
the implications for our L1157 model are as follows:
during Phase I of our model, when the temperature is 10K, the bulk of the ammonia is formed on the grains.
During Phase II, the bulk of the ammonia is released from the grain mantles 
due to sputtering at t$_{sat}$ $\sim$ 5 years (i.e at $z$ $\sim$ 5$\times$10$^{14}$ cm); however 
an increase in abundance (from 4 to 6$\times$10$^{-5}$) does occur during the passage of the shock. The increase
corresponds to 
when the temperature reaches $\sim$1200K (corresponding to v$_n$ $\sim$ 15 kms$^{-1}$), 
but it is short-lived as NH$_3$ starts 
declining when the temperature reaches 3500K (corresponding to v$_n$ $\sim$ 25 kms$^{-1}$). 
In terms of emission size of the clump this corresponds to less than 0.001 pc
(0.5$^{"}$--0.8$^{"}$ assuming a distance of 400 and 250 pc respectively).
This implies that observations of NH$_3$ 
toward the B1 clump with single dishes ($\sim$20$^{"}$--40$^{"}$) 
would then give lower NH$_3$ abundances than those predicted by our model 
due to beam dilution (i.e. 5$\times$10$^{-5}$ $\times$ [1$^{"}$/(20$^{"}$--40$^{"}$)$^2$] $\sim$ 10$^{-7}$--3$\times$10$^{-8}$). We note that these diluted abundances are similar
to those reported by Bachiller et al. (1993) toward the B1 clump. 
High-angular resolution observations of high-excitation lines of NH$_3$ 
with the VLA are therefore required to establish the length of 
the region within the postshock gas where NH$_3$ is expected to be enhanced.
\par
The degree and extent of 
the increase in the NH$_3$ abundance due to gas phase high temperature chemistry will also depend on the degree
of depletion adopted in Phase I; for example, if one compares Model 1 and Model 3 (where 1\% and 30\% of the gas phase 
is depleted by the end of Phase I, respectively) one notes that in Model 1 the relative increase is more substantial (at $\sim$ 3$\times$10$^{15}$ cm), due to the lower NH$_3$ abundance released from the grains in Phase II. 
\par Ultimately, therefore, the difference between the water and ammonia behaviours between
Models 1 and 3, and Models 4 and 7 is due to the temperature: in a model where the initial
density is $<$ 10$^5$ cm$^{-3}$, or the velocity is $<$ 40 kms$^{-1}$,  
the maximum temperature reached by the C-shock is
$\sim$ 3000 K. Under these conditions ammonia is maintained high as there is no effective destruction route while
it can efficiently form via the reaction H$_2$ + NH$_2$.
For models where the temperature during the shock is $\sim$ 4000 K, ammonia is instead efficiently destroyed by the 
reaction: H + NH$_3$ $\Rightarrow$ NH$_2$ + H$_2$. The latter 
has
a high barrier ($\sim$ 5000K) 
and hence it is inefficient for models where the maximum temperature of the shock is lower than 4000K.
The main reaction that forms water, on the other hand, is very efficient as long as temperatures are over 220K while the main destruction route (via reaction with atomic hydrogen) has a very high barrier ($\sim$ 9000K).  
\par 
As the discriminating factor seems to be the maximum temperature attained by the shock we note that if the shock velocity is higher (e.g. 60 kms$^{-1}$) then low initial 
densities may still lead to maximum temperatures of $\sim$ 4000K (Models 8-10). However we find that, regardless of the shock parameters, low initial densities imply that the gas abundances of water and ammonia (and other species) do 
not reach high values until the shock is at least half way through the dissipation region, leading to underestimated column densities. Moreover, 
if a shock velocity of 60 kms$^{-1}$ 
were present toward B1, we
should expect to detect H$_2$O at much higher terminal velocities than
observed ($\sim$30 kms$^{-1}$).
These qualitative comparisons indicate that the best fit dynamical model for L1157 is one where the shock velocity is $\sim$ 40 kms$^{-1}$ and 
n$_H$ $\sim$ 10$^5$ cm$^{-3}$.

\par 
Comparing our water abundances with those
found by Lefloch et al. (2010) we are in good qualitative agreement: 
they find
that the water is enhanced by two orders of magnitude between the low (from 
V$_{LSR}$=2.5 to -6 km s$^{-1}$) 
and high-velocity (from
V$_{LSR}$=-6 to -25 km s$^{-1}$) components from 8$\times$10$^{-7}$ to
8$\times$10$^{-5}$ respectively; comparing these abundances with our lowest pre-shock water
abundances of 8$\times$10$^{-7}$ (corresponding to V$_{LSR}$ up to -6 kms$^{-1}$) and the high abundance reached during
the shock of $\sim$ 2$\times$10$^{-4}$ (from -7 to -30 kms$^{-1}$), respectively; the latter is a factor of 3
higher than observed; 
we consider a match
within a factor of three as a good agreement.

\par Finally, we test whether the profile of the water line does indeed differ from that of ammonia.
We have used the velocity, abundance, temperature and density profiles from 
Models 3 and 4 
as input to our radiative transfer model SMMOL and we computed the line profiles for water and  ammonia. 
Figure 2 shows water and ammonia profiles for Model 3, overplotted with the ammonia profile for Model 4 (the water profile for Model 4 is very similar to that of Model 3).
Clearly, while certainly not a match, the line profile for water is different from that of ammonia for Model 3, and 
in particular it extends to higher velocities. Moreover, 
in very good agreement with the observations, the water intensity is about a factor of 7 higher than that of ammonia (the latter is
multiplied by this factor in the figure - see caption). On the contrary, the NH$_3$ profile for Model 4
extends at higher velocities and its intensity is higher than observed (in the figure it is multiplied by a factor of 5).
This is the first time
that modelling very high resolution profiles of water and ammonia 
in shocked regions starting from
a pure ab initio approach
has been attempted.
Note that we do not aim at a quantitative match
with the high resolution HIFI spectra: this would require a better knowledge of the geometry and structure of the B1 clump 
and most importantly a three dimensional radiative transfer model. Nevertheless this first attempt is very promising as we are in fact able to reproduce how the NH$_3$/H$_2$O intensity ratio changes with velocity. 


\begin{figure}
\includegraphics[angle=-90,width=18cm]{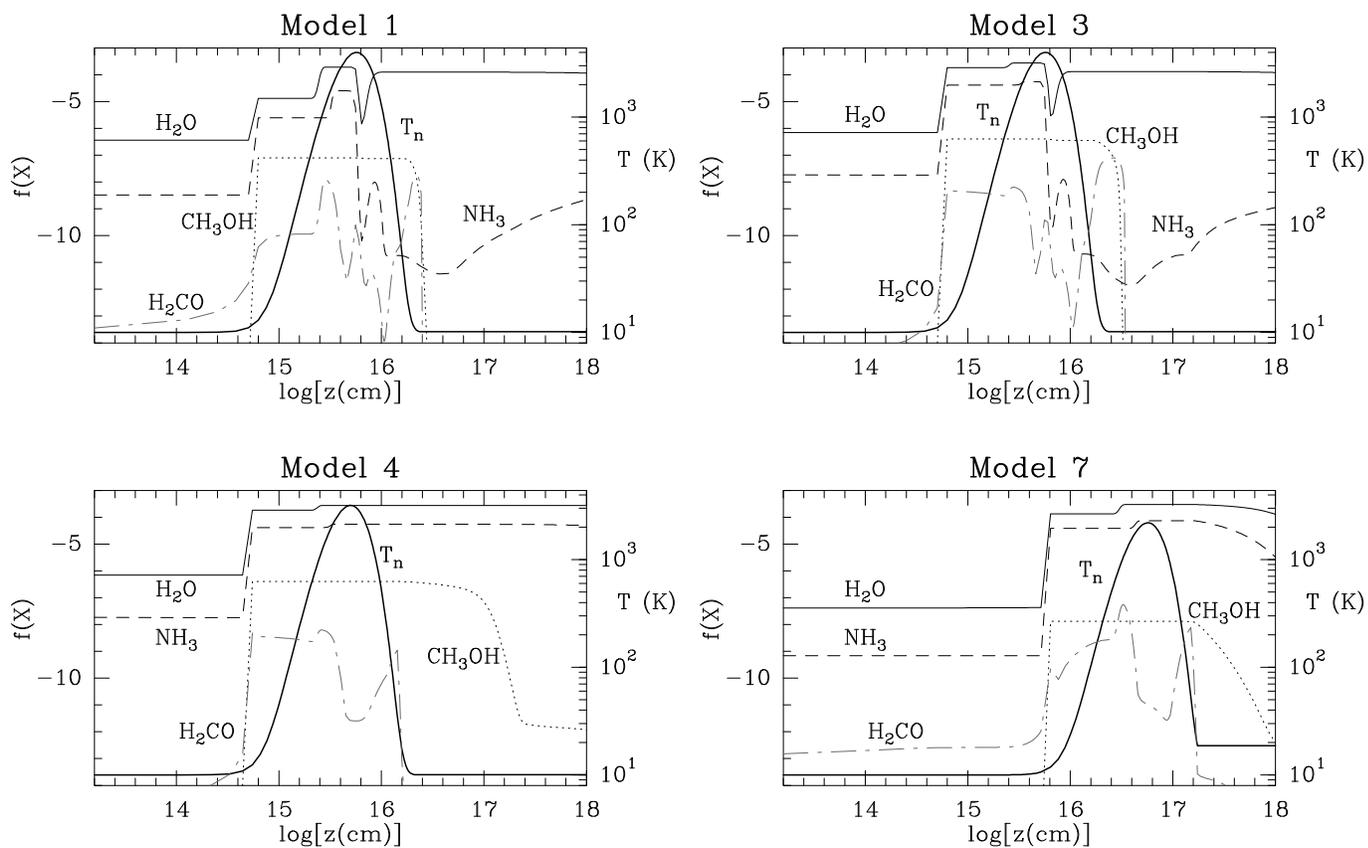}
\caption{Fractional abundances (with respect to molecular hydrogen),
of selected species as a function of distance for Phase II for Models 1, 3, 4 and 7. 
The temperature profile
for each model is also shown.}
\label{fig:abundance}
\end{figure}

\begin{figure}
\includegraphics[angle=-90,width=15cm]{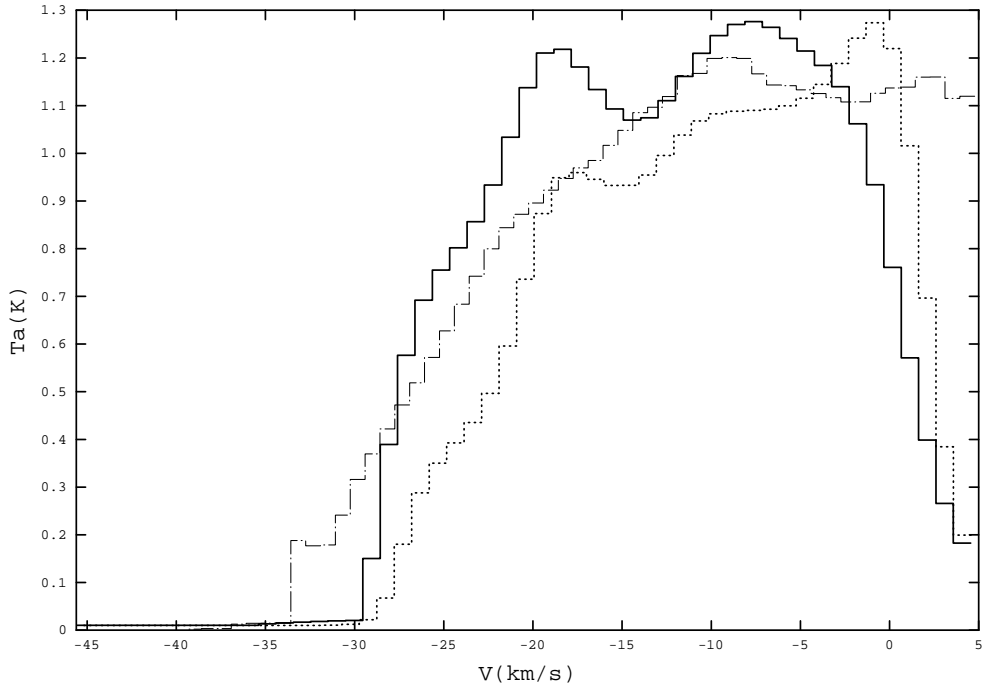}
\caption{Theoretical line profiles of H$_2$O(1$_{10}$--1$_{01}$) (solid line), and NH$_3$(1$_0$--0$_0$) (dotted line) for Model 3. The profile of 
the NH$_3$ line has been multiplied by a factor of 7.5. The V$_{LSR}$ is +2.6 kms$^{-1}$. We also overplot the theoretical line profile of NH$_3$(1$_0$--0$_0$) (dot-dash line) for Model 4, multiplied by a factor of 5.}
\label{fig:profile}
\end{figure}

\section*{Acknowledgements}
IJS acknowledges the Smithsonian Astrophysical Observatory for the
support provided through a SMA fellowship.

\vspace{1cm}
\noindent
{\bf References} \\ \\

{\small
\noindent
Bachiller R., R.; Mart\'{\i}nn-Pintado, J.; Fuente, A., 1993, ApJ, 417, L45 \\
\noindent
Bachiller R., \& Per\'ez Guti\'errez M. 1997, ApJ, 487, L93 \\
\noindent
Bachiller R., Per\'ez Guti\'errez M., Kumar M.S.N., \& Tafalla M. 2001, A\&A 372, 899 \\
\noindent
Bendettini, M, Yates, J A, Viti S, Codella C, 2006, MNRAS, 370, 229 \\
\noindent
Benedettini M., Viti, S.; Codella, C.; Bachiller, R.; Gueth, F.; Beltr\'an, M. T.; Dutrey, A.; Guilloteau, S., 2007, MNRAS, 381, 1127 \\
\noindent
Beuther H., Thorwirth, S., Zhang, Q., Hunter, T. R., Megeath, S. T., Walsh, A. J., Menten, K. M., 2005, ApJ, 627, 834  \\
\noindent
Brown P. D., Charnley, S. B., Millar, T. J., 1988, MNRAS, 231, 409 \\
\noindent
Caselli, P., Hartquist, T. W., \& Havnes, O. 1997, \aap, 322, 296 \\
\noindent
Ceccarelli C., Bacmann A., Boogert A., et al., 2010, A\&A, 521, L22 \\
\noindent
Chieze, J.-P.; Pineau des For\^ets, G.; Flower, D. R., 1998, MNRAS, 295, 672 \\
\noindent
Codella C., Benedettini M., Beltr\'an M.T., et al. 2009, A\&A, 507, L25 \\
\noindent
Codella C., Lefloch B., Ceccarelli, C., et al. 2010, A\&A, 518, L112 \\
\noindent
Danby G., Flower D. R., Valiron P. Shilke P., 1988, MNRAS 235, 229  \\
\noindent
Draine  B. T.; Roberge, W. G.; Dalgarno, A., 1983, ApJ, 264, 485 \\
\noindent
Faure A, Crimier N, Ceccarelli C, Valiron P, Wiesenfeld L, Dubernet M L, 2007, A\&A,472, 1029 \\
\noindent
Flower, D. R., \& Pineau des For\^ets, G. 2003, \mnras, 343, 390 \\
\noindent
Gueth F., Guilloteau S., \& Bachiller R. 1996, A\&A 307, 891 \\
\noindent
Ho P T P, Townes C H, 1983, ARA\&A, 21, 239 \\
\noindent
Jim\'enez-Serra, I., Caselli, P., Mart\'{\i}n-Pintado, J., \& Hartquist, T. W.
2008, \aap, 482, 549 \\
\noindent
Jim\'enez-Serra, I.;
Mart\'{\i}n-Pintado, J., Rodr\'{\i}guez-Franco, A., \& Mart\'{\i}n, S.
2005, ApJ, 627, L121 \\
\noindent
Kaufman, M. J., \& Neufeld, D. A. 1996, \apj, 456, 250 \\
\noindent
Garay G., K\"ohnenkamp I., Bourke T.L., Rodr\'{\i}guez L.F., \& Lehtinen K.K. 1998, ApJ, 509, 768 \\
\noindent
Gusdorf A., Pineau Des For\^ets G., Cabrit S., \& Flower D.R. 2008, A\&A 490, 695 \\
\noindent
J\o{}rgensen J.K., Bourke T.L., \& Myers P.C. 2007, ApJ, 659, 479 \\
\noindent
Le Bourlot, J.; Pineau des ForȬts, G.; Flower, D. R.; Cabrit, S., 2002, MNRAS, 332, 985 \\
\noindent
Lefloch B., Cabrit S., Codella C., et al. 2010, A\&A, 518, L113 \\
\noindent
Lerate, M. R.; Yates, J. A.; Barlow, M. J.; Viti, S.; Swinyard, B. M., 2010, MNRAS, 406, 2445 \\
\noindent
Lesaffre P.; Chi\'eze, J.-P.; Cabrit, S.; Pineau des For\^ets, G, 2004, A\&A, 427, 147L \\
\noindent
Maret, S.; Faure, A.; Scifoni, E.; Wiesenfeld, L., 2009, MNRAS, 399, 425 \\
\noindent
May, P. W., Pineau des For\^ets, G., Flower, D. R., Field, D.,
Allan, N. L., \& Purton, J. A. 2000, \mnras, 318, 809 \\
\noindent
Neufeld D A; Nisini, B; Giannini, T; Melnick, G J.; Bergin, E A; Yuan, Y; Maret, S.; Tolls, R.; G\"usten,lf; Kaufman, M J., 2009, ApJ, 706, 170 \\
\noindent
Nisini, B, Giannini, T, Neufeld, D, Yuan, Y, Antoniucci, S, Bergin, E A, Melnick, G, 2010, ApJ, 724, 69
\\
\noindent
Rawlings J. M. C.; Yates, J. A., 2001, MNRAS, 326, 1423 \\
\noindent
Rawlings J. M. C.; Hartquist, T. W.; Menten, K. M.; Williams, D. A., 1992, MNRAS, 255, 471 \\
\noindent
Rybicki G. B., Hummer D. G., 1991, A\&A, 245, 171 \\
\noindent
Sepulveda I., Anglada, G., Estalella, R., L\'opez, R.; Girart, J. M., Yang, J., 2011, A\&A, 527, 41\\
\noindent
Tafalla M., \& Bachiller R. 1995, ApJ 443, L37 \\
\noindent
van Loo, S.; Ashmore, I.; Caselli, P.; Falle, S. A. E. G.; Hartquist, T. W., 2009, MNRAS, 395, 319 \\
\noindent
Viti, S, Collings, M P, Dever, J W, McCoustra M R S, Williams D A, 2004, MNRAS, 354, 1141 \\
\noindent
Viti, S, Codella, C, Benedettini M, Bachiller R, 2004, MNRAS, 350, 1029 \\
Zhang Q., Ho T. P., Wright M. C. H., Wilner D. J., 1995, ApJ, 451, L71
}
%
\end{document}